\documentclass[twocolumn]{article}

\usepackage{amsmath,amssymb,amsfonts}
\usepackage{balance}
\usepackage{siunitx}
\usepackage{graphicx}
\usepackage{bm}
\usepackage{gensymb}
\usepackage{float}
\usepackage[dvipsnames]{xcolor}
\usepackage{authblk}  
\usepackage{hyperref}  

\usepackage[round,authoryear,sort&compress]{natbib}
\usepackage{har2nat}
\begin{document}

\title{\textbf{Characterization of Supersonic Jet and Shock Wave with High-Resolution Quantitative Schlieren Imaging}}

\author[1,2]{Yung-Kun Liu\thanks{r06222017@ntu.edu.tw}}
\author[1,2]{Ching-En Lin}
\author[1,2]{Jiwoo Nam}
\author[1,2]{Pisin Chen\thanks{pisinchen@phys.ntu.edu.tw}}

\affil[1]{Leung Center for Cosmology and Particle Astrophysics, National Taiwan University, Taipei 10617, Taiwan, R.O.C.}
\affil[2]{Department of Physics, National Taiwan University, Taipei 10617, Taiwan, R.O.C.}

\renewcommand\Affilfont{\itshape\small}

\date{}
\maketitle

\begin{abstract}
This paper presents an enhanced optical configuration for a single-pass quantitative Schlieren imaging system that achieves an optical resolution of approximately 4.6 $\mu m$. The modified setup decouples sensitivity from resolution, enabling independent optimization of these critical parameters. Using this high-resolution system, we conduct quantitative analyses of supersonic jets emitted from sub-millimeter nozzles into the atmosphere and investigate shock waves induced by knife blades interacting with these jets in a vacuum environment. The fine resolution allows for detailed visualization of shock wave structures and accurate measurement of density gradients. We demonstrate the system's effectiveness by examining the density gradient profile along the shock diamonds and mapping density profiles across shock waves. These density profiles are analyzed for their relevance in laser-plasma applications, including laser wakefield acceleration and the Analog Black Hole Evaporation via Laser (AnaBHEL) experiment. Our findings indicate that this system can help determine key parameters such as peak density, plateau length, and shock wave thickness—essential for optimizing electron acceleration and achieving specific plasma density profiles. This high-resolution quantitative Schlieren imaging technique thus serves as a valuable tool for exploring complex fluid dynamics and supporting advancements in laser-plasma physics research.
\end{abstract}


\section{Introduction}
Optical techniques such as Schlieren imaging, interferometry, and shadowgraph are pivotal for determining the refractive index in transparent targets and are extensively utilized in fluid dynamics studies. These techniques differ fundamentally: interferometry measures the phase difference accumulated between the signal and the reference beam, Schlieren imaging detects refractive index gradients, and shadowgraphy captures the second derivative of the refractive index. Consequently, they provide measurement of density ($\rho$), its first derivative ($\partial \rho$), and its second derivative ($\partial^2 \rho$), respectively. For an in-depth discussion of these methods, see \citep{settles-2001}. 

While each of these methods has its strengths, Schlieren imaging offers distinct advantages, particularly in the study of supersonic flows and shock waves. Its relatively simple setup, combined with its natural suitability for characterizing strong discontinuities in fluid properties, makes it an invaluable tool in many research areas. A typical Schlieren system comprises a light source, optics, a target section, a cut-off, and a sensor. The cut-off, often a simple knife blade placed at the focal point of the first lens after the target, is crucial as it modulates the system's response to density gradients.

The versatility and precision of optical diagnostic techniques have led to their application in various advanced fields, including plasma wakefield acceleration (PWA). PWA technologies, such as laser wakefield acceleration (LWFA) \citep{tajima1979laser} and particle wakefield acceleration (PWFA) \citep{chen1985acceleration}, have garnered significant attention due to their ability to generate electric fields on the order of tens of GeV/m, nearly three orders of magnitude above traditional RF accelerators. These technologies hold the promise of more compact and efficient particle accelerators for applications ranging from high-energy physics to medical treatments.

A critical factor in the performance of WFA technologies is the precise control and characterization of the plasma density profile. The quality of the accelerated electron beam, particularly its energy spread and emittance, is highly sensitive to the plasma density distribution. Any variations or inhomogeneities in the density profile can lead to beam quality degradation, premature dephasing, or reduced acceleration efficiency. Furthermore, the injection process, where background electrons are trapped and accelerated by the wakefield, is strongly influenced by the density gradient. Controlled injection techniques, such as density downramp injection \citep{bulanov1998injection}, rely on precise shaping of the longitudinal density profile to achieve high-quality, mono-energetic electron beams. Consequently, accurate characterization of supersonic gas jets, which are commonly used to create the required plasma densities, is of paramount importance for optimizing PWA performance.

Furthermore, in 2017, Chen and Mourou proposed the Analog Black Hole Experiment via Laser (AnaBHEL) \citep{pisin2017AnaBHEL,pisin2020mirrorTrajectory,pisin2022anabhel}, integrating concepts from laser wakefield acceleration and the flying mirror model from quantum field theory in curved spacetime. In this groundbreaking experiment, the density profile and its gradient are crucial for determining the temperature of the analog black hole, further emphasizing the need for high-resolution imaging techniques.

To address these evolving requirements, this paper introduces an advanced single-pass Schlieren imaging system with an optical resolution of approximately $4.6\mu m$. Our system not only achieves this fine resolution but also provides a degree of independent adjustment between sensitivity and resolution. This improvement allows for more detailed capture of shock wave structures and precise characterization of density profiles in supersonic jets emitted from sub-millimeter nozzles.

While other techniques such as interferometry or shadowgraphy \citep{golovin2016shadow,hansen2018interferometry,kim2018shadow}, tomography \citep{couperus2016tomographic,adelmann2018tomography}, Planar Laser-Induced Fluorescence (PLIF) \citep{mao2017gas,hanson2018handbook} and Rayleigh scattering \citep{KREISLER_1980_RayleightScattering} have been explored for similar applications, the Schlieren system we proposed in this paper offers a balance of high resolution, sensitivity, and relative simplicity that makes it particularly suitable for studying supersonic jets and shock waves.

The paper is organized as follows: Section 2 outlines the fundamental principles of Schlieren imaging and various system configurations. Section 3 details our experimental setup, including the single-pass Schlieren system and the gaseous target. Section 4 discusses the experimental results and analysis, focusing on the system's resolution, factors influencing image quality, and the application in determining density profile across the shock wave. The potential applications in WFA and AnaBHEL experiments are also addressed.

\section{Overview of Schlieren imaging principles}
\subsection{Principles}
Schlieren imaging is a powerful optical technique used to visualize and measure density gradients in transparent media. This method is based on the principle that light rays are deflected when passing through regions with varying refractive indices. Consider a collimated light beam traveling in the x-direction through a medium with a refractive index $ n = n(x,y,z) $, the deflection angle $ \epsilon_y $ in the y-direction is given by:
\begin{equation}
    \epsilon_y (y,z)=\frac{1}{n_0}\int_{0}^{L}\frac{\partial n (x,y,z)}{\partial y}dx
    \label{eq:deflection_angle_Schlieren_principle}
\end{equation},
where $ 0 \le x \le L $ is the extent of the target and $ n_0 $ is the ambient refractive index. An analogous expression applies for deflections in the y-direction. For gaseous targets, the refractive index can be related to the density using the Gladstone-Dale relationship:
\begin{equation}
    n-1=\kappa\rho
    \label{eq:Gladstone-Dale_relation}
\end{equation}
, where $ \kappa $ represents the Gladstone-Dale coefficient, approximately $ 2.38 \times 10^{-4} \, \text{m}^3/\text{kg} $ for air. Therefore, the density gradient of the target can be deduced from the deflection angle measured via Schlieren techniques. Equation(\ref{eq:deflection_angle_Schlieren_principle}) shows that the measured deflection angle is an integration over the line of sight. To retrieve the refractive index distribution, certain assumptions must be made. For example, in the case of a conical nozzle with axial symmetry, the density profile of the jet can be assumed to be axisymmetric, allowing the use of an inverse Abel transformation to solve $\partial n(y,z)/\partial y$ from $\epsilon_y(y,z)$. On the other hand, if the refractive index distribution is nearly uniform along the line of sight, such as in a jet from a slit nozzle, the integration can be further approximated. A slit nozzle is a type of nozzle characterized by its elongated, narrow opening, as opposed to a circular or conical exit. This design produces a jet with a more uniform, two-dimensional flow pattern in comparison to the axisymmetric jets typically produced by circular nozzles. In this case, eq.(\ref{eq:deflection_angle_Schlieren_principle}) can be written into \begin{equation}
    \epsilon_y (y,z)\approx\frac{L}{n_0}\frac{\partial n (x,y,z)}{\partial y}
    \label{eq:deflection_angle_Schlieren_principle_simplified}
\end{equation}. By combining Eqs.(\ref{eq:Gladstone-Dale_relation}) and (\ref{eq:deflection_angle_Schlieren_principle_simplified}), the density gradient can be approximated as: 
\begin{equation}
\frac{\partial \rho(y,z)}{\partial y}\approx \frac{n_0}{L\kappa}\epsilon_y (y,z)    \label{eq:simplified_density_from_deflection}
\end{equation}.

 \subsection{Different Schlieren Configurations}
The essence of Schlieren imaging lies in measuring the light deflection $\epsilon$, typically achieved by allowing the light to propagate for a certain distance and measuring the displacement relative to the unperturbed light. Various methods have been proposed to characterize this deflection, a comprehensive review can be found in the textbook \citep{settles-2001} and the review paper \citep{ComparisonSchlieren_Hargather2012,settles2017review}. One common way is utilizing a cut-off, which is usually placed at the focal point of the first lens after the target. If the focal length is $f_1$, the displacement of the focal point at the cut-off plane for a light ray deflected by an angle $\epsilon_y$ is given by
\begin{equation}
    \Delta y = L_{equiv}\times\epsilon_y
    \label{eq:light_displacment}
\end{equation}
, where $L_{equiv}$ is the equivalent propagation distance of the light from the target. For a single-pass Schlieren imaging system, $\Delta y = f_1\times\epsilon_y$ hence $L_{equiv}=f_1$. This distance determines the sensitivity of the Schlieren system; a larger $L_{equiv}$ induces a greater displacement, making it easier to observe.

\begin{figure}[H]
     \centering
     \includegraphics[width=\linewidth]{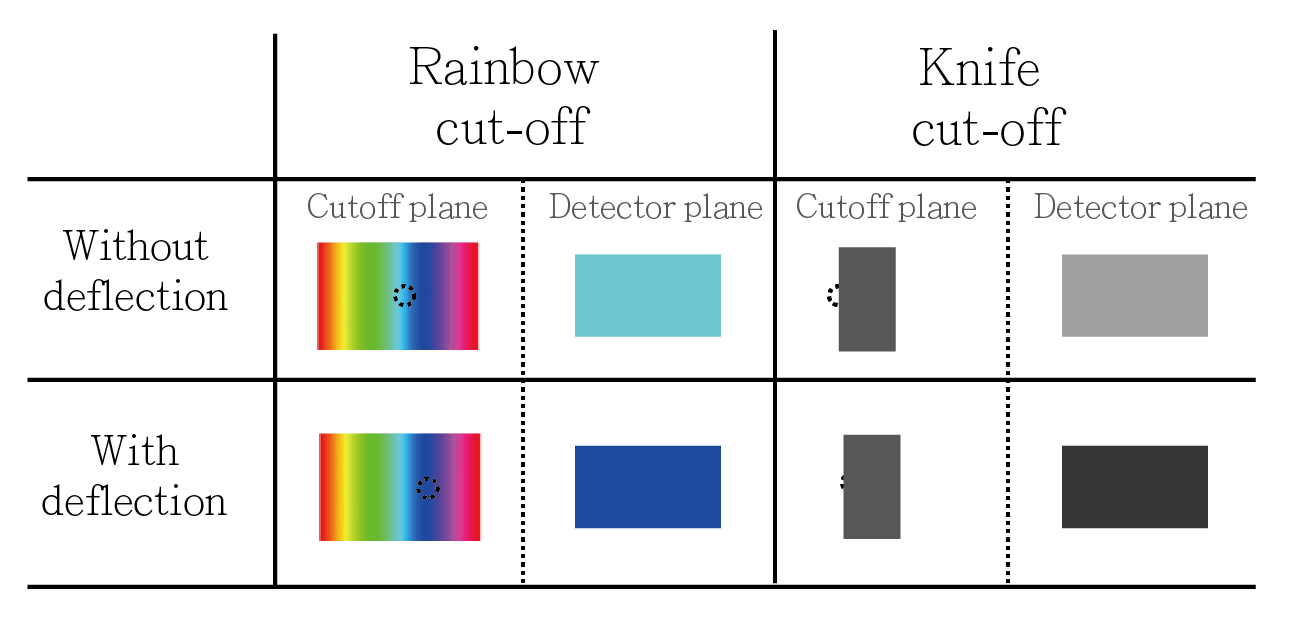}
     \caption{Schematic comparing rainbow and blade cut-offs in Schlieren imaging. The black dashed circle represents the focused light at the cut-off plane. The detector plane shows the corresponding image as seen by the camera. For the rainbow cut-off, color changes indicate light deflection, while for the blade cut-off, brightness variation is caused by partial light obstruction.}
     \label{fig:schematic_different_cutOff}
 \end{figure}

Different cut-offs vary in how they record the displacement $\Delta y$. Here, we briefly discuss two widely used cut-offs (the schematic is shown in Figure.\ref{fig:schematic_different_cutOff}):
 \begin{itemize}
     \item Blade cut-off: The blade cut-off is extensively used in qualitative Schlieren imaging, due to its easy preparation: just a simple razor blade. It's worth mentioning that a blade cut-off can also be used for quantitative measurement. This relies on the fact that the light source is finite in size. Consider at the cut-off plane, the uncovered beam (assumed to be a square shape for simplicity) height is $a$. When there is a displacement $\Delta y$ due to the target density gradient, different percentages of the light will be blocked by the cut-off, which results in a change in the detected brightness. In the above example, the brightness variation is $\frac{\Delta E}{E}=\frac{f_1\epsilon_y}{a}$. The deflection can then be deduced from the brightness measurement. The blade cut-off can offer high sensitivity and is particularly effective for visualizing small-density gradients.
     \item Rainbow filter: Different from using the change of brightness in the blade cut-off case, the idea is to put a colored filter with a designed pattern, for example, a ring shape \citep{colorSchlieren_Kleine1991,RingSchlierenStevenson2015} or rainbow stripes in \citep{RainbowSchlieren_tomography2018,Mariani2019,Mariani2020}. The light displacement is therefore encoded in the color (typically the Hue). Using Hue to encode the signal brings another advantage: less sensitivity to the background light disturbance. With different designs of the color filter, a different sensitivity or detection range can be realized, which provides quite high flexibility for the measurement. 
 \end{itemize}
The orientation of the cut-off determines the measured density gradient direction. As an example, if a blade cut-off is put vertically (along the z-axis), the horizontal (in the y direction) light deflection can be distinguished, which corresponds to a measurement of $\partial n/\partial y$. Depending on the purpose, different orientations, or in a more general word, different symmetric axes, of the cut-offs can be designed to fit the requirement. In some cases, radial symmetric cut-offs can be designed to measure the density gradient along a radial direction.

\begin{figure}[H]
     \centering
     \includegraphics[width=\linewidth]{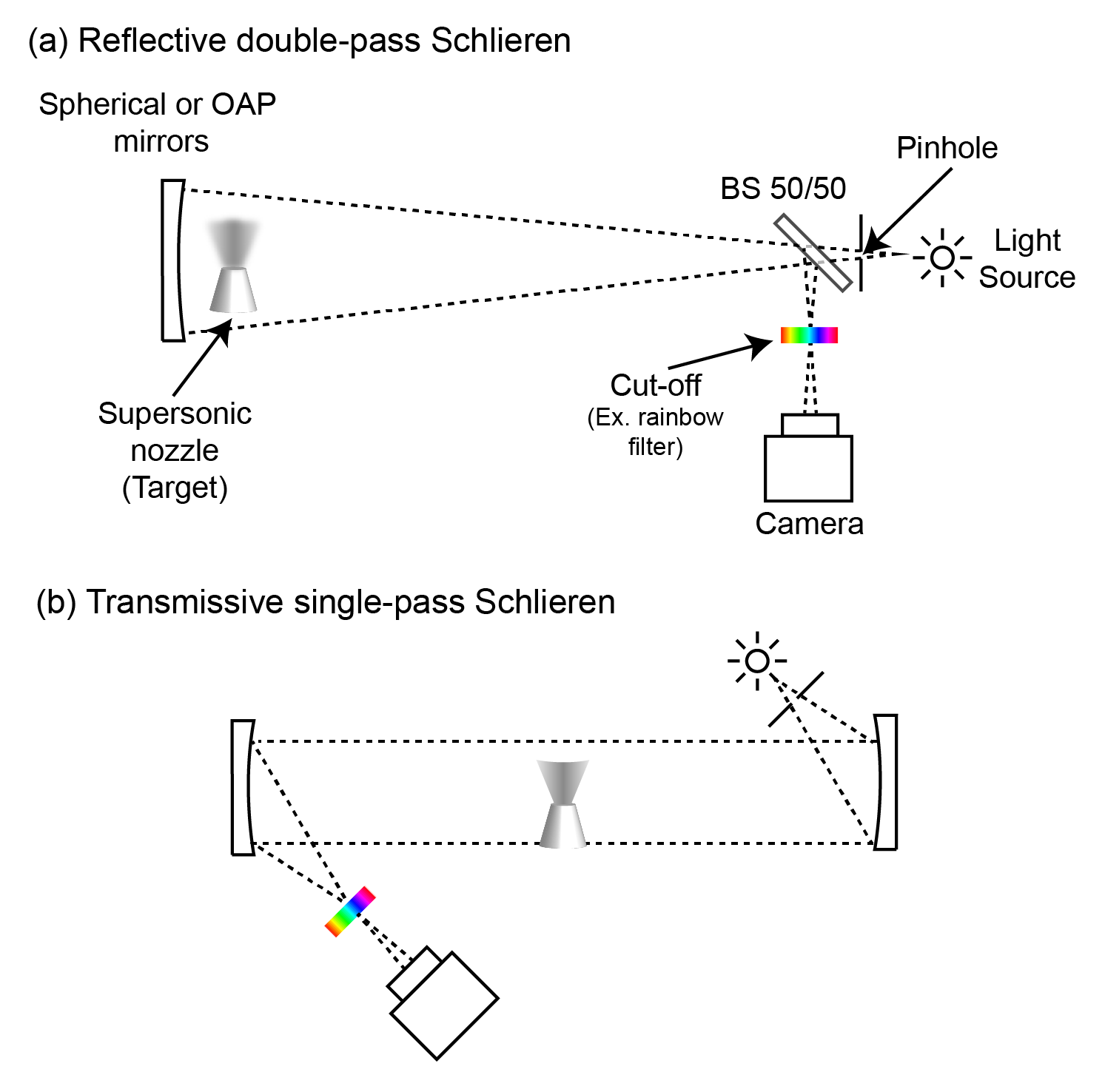}
     \caption{Schematic comparison of a double-pass Schlieren imaging system and a Z-type single-pass Schlieren system, illustrating their different optical configurations.}
     \label{fig:single-double-pass-Schlieren}
 \end{figure}

Besides the type of cut-off, different configurations of the Schlieren imaging can also be found in the literature. We may divide them into two groups based on how many times the light passes through the target, which can be a single-pass or double-pass setup. Figure.\ref{fig:single-double-pass-Schlieren} shows the schematic of a common double-pass and a Z-type single-pass Schlieren system. The advantage of the double-pass system is that the deflection angle accumulates to twice that of the single-pass system, promising to reach higher sensitivity; Besides, by using reflective optics and a beam splitter, the system has an advantage in an easier alignment process, which makes the double-pass system is suitable for most applications. An example setup, Figure.\ref{fig:single-double-pass-Schlieren} (a), contains a point light source (can be an LED with lenses and a pinhole ), a 50/50 beam splitter, a spherical mirror that is located twice its focal length away from the pinhole, a cut-off, and the camera. On the other hand, a common Z-type single-pass setup is shown in Figure.\ref{fig:single-double-pass-Schlieren} (b) with a similar setup but the light only passes once through the target. Each setup has various deformations, which can be found in \citep{settles-2001}. 

In our implementation, we found that though the double-pass imaging system enjoys higher sensitivity when taking images of the supersonic jet from our sub-mm nozzle, the results will suffer from edge diffraction which originates from the nature of the double-pass configuration: the light passes through the target twice, there are equivalently two targets, one is the real target, the other is the virtual image from the reflective optics. We can only focus on one of the target planes, which makes the other generate the edge diffraction pattern. This effect is negligible when the resolution requirement is much more coarse than the edge diffraction spacing, which is typically around a few to tens of $\mu m$. However, for the imaging system aimed to reach $\mu m$ optical resolution, this issue should be overcome. 

This limitation led us to explore a single-pass setup for our high-resolution imaging needs. We experimented with both rainbow filter and blade cut-off options. Using the USAF 1951 resolution target, we discovered that the rainbow filter limited our system's resolution to a minimum value of around 20 $\mu m$. In contrast, the blade cut-off had minimal impact on the resolution while still providing adequate sensitivity for our purposes. This combination of a single-pass setup with a blade cut-off allowed us to achieve the high resolution required for our study of supersonic jets from sub-mm nozzles while avoiding the edge diffraction issues associated with double-pass systems. Detailed results of our system's performance will be presented in the Results section.

\section{Experimental setup}
\begin{figure}[H]
     \centering
     \includegraphics[width=\linewidth]{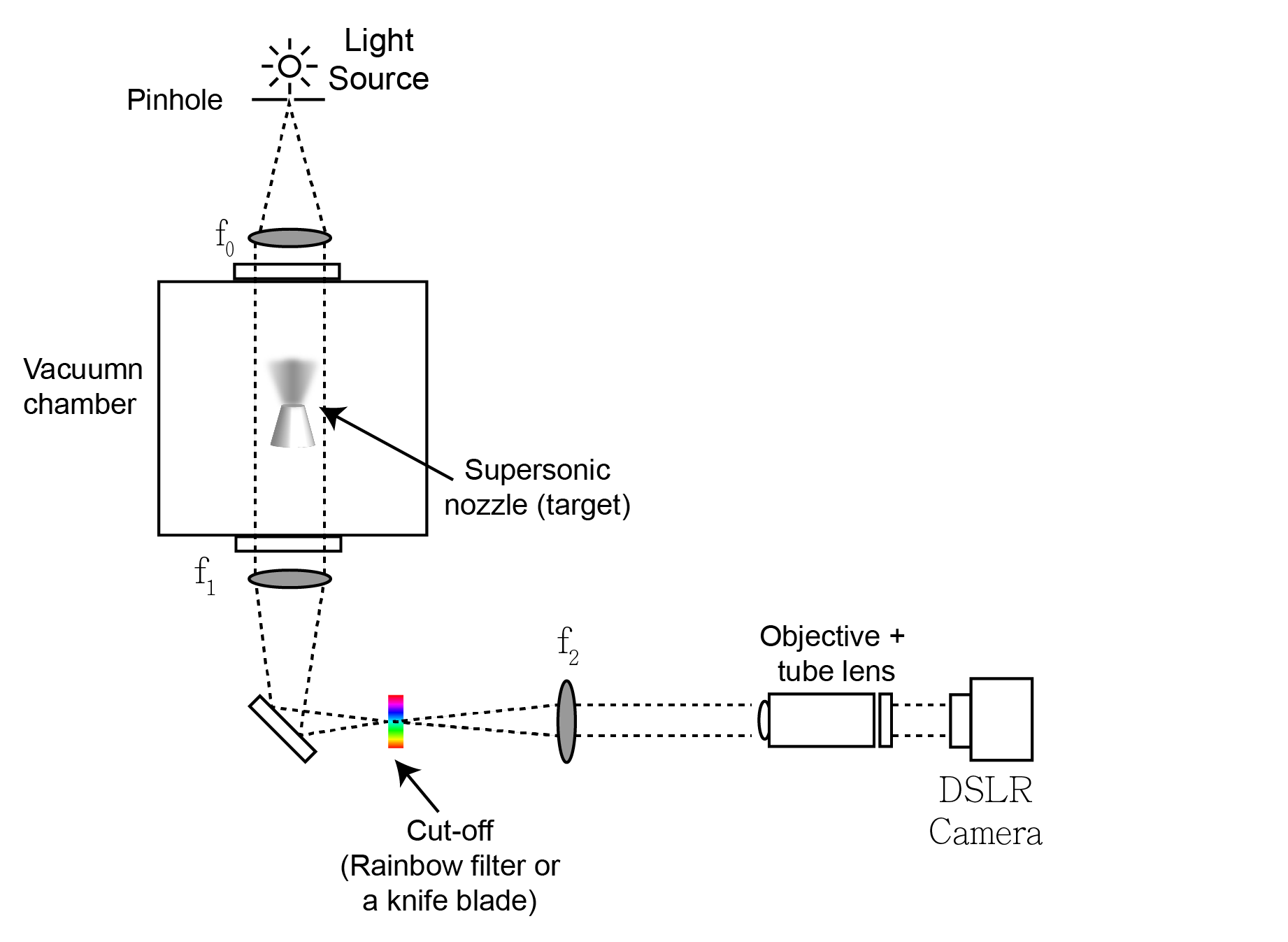}
     \caption{Schematic representation of the experimental setup. Detailed descriptions of each component are provided in the text.}
     \label{fig:schematic_experiement_setup}
 \end{figure}

Our experimental arrangement comprises two primary components: the Schlieren imaging system and the gaseous target. The imaging system consists of a 60W white light LED, collimating and imaging optics, a cutoff, and a detector. The gaseous target is a supersonic jet emitted from a custom sub-millimeter nozzle, supplied with 5N Argon from a high-pressure tank equipped with a pressure regulator and gauge. To facilitate the study of supersonic jet behaviors in a vacuum environment, the gaseous target is encased within a small vacuum chamber. Figure. \ref{fig:schematic_experiement_setup} presents a schematic of the experimental setup.

\subsection{Light source and detector}
The selection of an appropriate light source for Schlieren imaging involves considerations of brightness, coherence, and spectrum. Brightness influences the minimum achievable shutter speed, which is essential for capturing transient fluid dynamics. Typically, an incoherent light source is preferred to avoid speckle patterns that degrade image quality. The choice of the spectrum depends on the cutoff type; a broad-spectrum source is necessary when using a rainbow filter, while a monochromatic source suffices for a knife cut-off Schlieren.

Recent advancements in light-emitting diode (LED) technology have made it possible to find white light LEDs with brightness comparable to Tungsten or Xenon lamps, but at a significantly lower cost. However, LEDs have the drawback of non-uniform brightness, which depends on the shape of the LED dies and the collimating optics design of the LED module. Our setup utilizes a 60W white light LED, paired with a dual achromatic lens pair (f=50mm and f=75mm) and a $\phi=100\mu m$ pinhole to create a uniformly illuminated point source.

The choice of detector depends on the required shutter speeds, spectral responses, and pixel size. In most cases, a commercial digital single-lens reflex camera (DSLR) can meet these requirements. We employ a Canon 90D DSLR in our experiment, which provides a minimum shutter speed of 1/8000 sec and a pixel size of $3.2 \mu m\times 3.2\mu m$. For the results demonstrated in this study, we selected a shutter speed of 10ms, balancing the ability to capture the dynamics of our supersonic jet of interest with adequate image brightness.

\subsection{Imaging setup}
\begin{figure*}[h]
     \centering
     \includegraphics[width=0.8\textwidth]{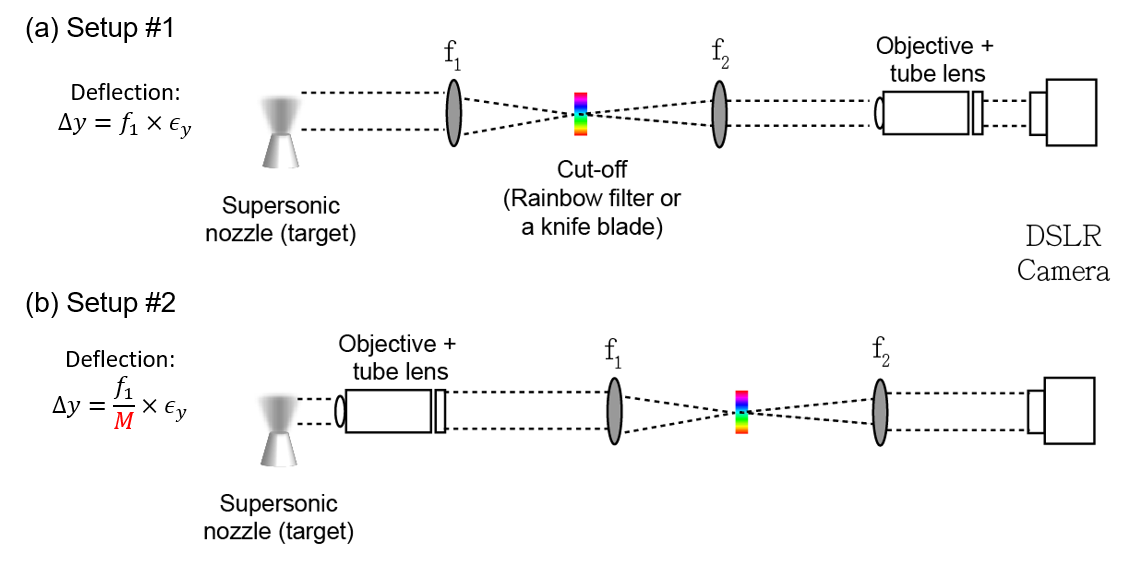}
     \caption{Two variations of the proposed decoupled Schlieren imaging system. In setup \#1, light with a deflection angle $\epsilon_y$ experiences $M$ times greater deflection at the cutoff plane compared to setup \#2, resulting in higher sensitivity. Although both setups have the same magnification ratio, the numerical aperture (N.A.) is limited by the relay lens in setup \#1 and by the objective lens in setup \#2, giving setup \#2 finer spatial resolution. The choice of setup depends on the specific application requirements.}
     \label{fig:schematic_decouple_two_variations}
 \end{figure*}

Figure \ref{fig:schematic_decouple_two_variations}a illustrates the schematic of our imaging setup. The imaging optics include two 2" achromatic lenses with focal lengths $f_1=f_2=300mm$, which function as an image relay to produce a 1:1 image at the intermediate image plane. This is followed by a 5X objective (Mitutoyo Plan APO long WD objective) and a tube lens (Mitutoyo MT-4) to produce a 5x magnified image on the camera. The cutoff is positioned at the focal point of the first achromatic lens.

In this configuration, the system's sensitivity depends on the choice of $f_1$. A longer focal length allows light to propagate further before reaching the cutoff, resulting in a larger displacement in the cutoff plane. Consequently, a larger $f_1$ provides higher sensitivity, which scales linearly with the focal length. The system's resolution, on the other hand, is primarily determined by the choice of objective and tube lens. Although this configuration is more complex than a typical single-pass Schlieren system, it offers the flexibility to independently adjust the sensitivity and magnification of the imaging system. In our design, we set $L_{\text{equiv}} = f_1 = 300$ mm (cf. Eq.~(\ref{eq:light_displacement})), providing sufficient sensitivity to capture the structure of the supersonic jet and shock waves. The resolution is approximately $4.6 \mu$m, as determined using a USAF 1951 resolution target.

It is important to note that sensitivity and resolution are not completely decoupled but have an implicit relationship when approaching the diffraction limit. The small numerical aperture (N.A.) of a long focal length lens may limit the achievable resolution. For instance, while a 5X objective with an N.A. of 0.14 has a resolving power of $2\mu m$, a $f=300$ mm plano-convex lens has an N.A. of approximately 0.085, which limits the overall resolution. This explains why our imaging system's resolution is $4.6\mu m$ rather than the $2\mu m$ specified for the objective.

A variation of this configuration places the imaging components (i.e., the objective and tube lens) in front of the relay optics, as shown in Figure \ref{fig:schematic_decouple_two_variations}b. Ray optics calculations show that in this case, the light ray displacement is $\Delta y = f_1 \times \epsilon_y / M$. The presence of magnification in the denominator indicates that this setup is five times less sensitive compared to the original configuration, even when using the same optical components. Although the magnification remains the same, in this setup the numerical aperture is limited by the objective. For example, with our chosen optics in setup \#2, the N.A. is 0.14, providing a theoretical resolving power of around $2 \mu$m.

In conclusion, there is no definitive choice between these two setups; rather, there is a trade-off. If the phenomenon has a mild density gradient requiring higher sensitivity, or if space around the target is limited, the first setup is preferable. Conversely, if finer resolution is essential, the second setup is a better choice.

\subsection{Cutoffs and calibrations}
We tested two different cutoffs in our experiment: a rainbow filter and a blade cutoff. The rainbow filter design is based on the work of Mariani et al. \citep{Mariani2019, Mariani2020}, featuring a linear variation of Hue from 0 to 1. Rainbow filters can be manufactured as positive films or printed using high-resolution printers on transparent film. In our case, we adopted the positive film development technique to produce filters with varying widths\footnote{As digital photography gradually replaces film photography, it is becoming increasingly challenging to find services offering positive film development.}. For the results presented in this paper using a rainbow filter as the cutoff, the filter was designed with Hue varying linearly from 0 to 1 within a width of 6mm. 

\begin{figure*}[t]
     \centering
     \includegraphics[width=0.8\textwidth]{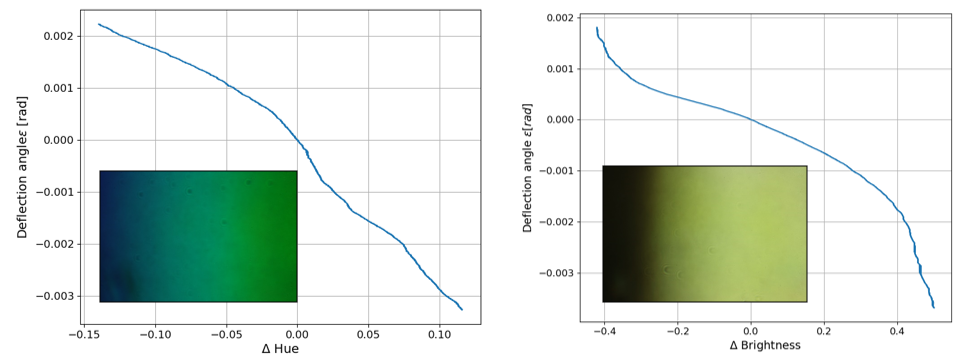}
     \caption{Calibration curves for the Schlieren imaging system with two different cutoffs: a 6mm rainbow filter and a knife blade. Inset: Schlieren image of a calibration lens with $f=1.0m$.}
     \label{fig:calibration_curve_different_cutoff}
 \end{figure*}

In quantitative Schlieren techniques, calibration is a crucial step that links the detected signal variation (which can be a difference in color (Hue) or brightness) to the refractive gradient. A typical approach involves using an object with a known refractive index distribution, such as a lens, as the target. Figure. \ref{fig:calibration_curve_different_cutoff} shows the calibration curve using a $f=1m$ plano-convex lens from Edmund Optics\footnote{The choice of focal length for the calibration lens depends on the magnification and sensitivity of the imaging system. For example, Mariani et al. \citep{Mariani2020} used an f=10m lens for calibration in a system with magnification less than one.}. The inset of the figure displays the Schlieren image, where a clear Hue/brightness variation across the lens is visible. Since the deflection angle of light after passing through the lens can be described by$\epsilon_r = r/f$, we can directly relate the deflection angle to the detected signal.

\subsection{Gaseous target}
The supersonic jet was generated using a cylindrical nozzle with a diameter of $500\mu m$, supplied with Argon gas at 115 psi. A pulsed valve with an opening time of 100ms controls the gas flow, reaching a steady state during the camera's exposure time. The nozzle is housed within a cubic vacuum chamber with a side length of 20 cm. In this study, "vacuum" refers to a pressure of 0.1 mbar, achieved using a rotary pump. Under these conditions, the supersonic jet is under-expanded, meaning the pressure at the nozzle exit is higher than the ambient pressure. Consequently, further decreases in ambient pressure do not significantly affect the supersonic jet.

To investigate blade-induced shock waves, we installed a customized stainless steel knife blade on a two-axis piezo-motorized stage (Thorlabs PD1/M) as an obstacle. The blade can move in the y and z directions, enabling control over both its coverage and height relative to the nozzle.

\section{Results}
\subsection{System resolution}
\begin{figure*}[t]
     \centering
     \includegraphics[width=\textwidth]{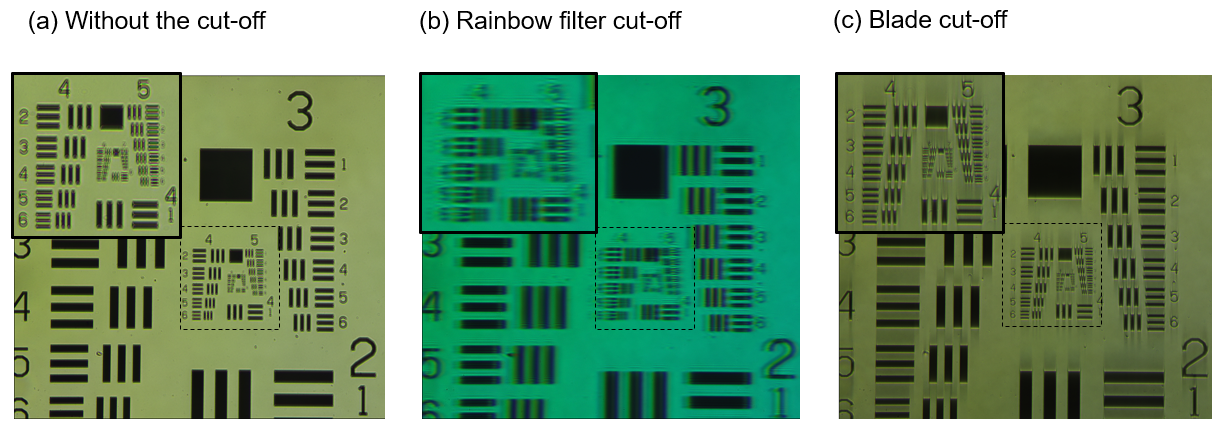}
     \caption{Images of a 1951 USAF resolution target captured by our Schlieren imaging system to determine the resolution under different conditions: (a) without cutoff, (b) with a 6mm rainbow filter, and (c) with a blade cutoff.}
     \label{fig:resolution_target_w_wo_cut-off}
 \end{figure*}

We employed the standard USAF 1951 resolution target to determine the resolution of our imaging system, which is defined by the finest patterns our system can resolve. Fig. \ref{fig:resolution_target_w_wo_cut-off} presents images of the resolution target captured using our decoupled single-pass Schlieren system. Fig. \ref{fig:resolution_target_w_wo_cut-off}a shows the result without a cutoff, while fig.\ref{fig:resolution_target_w_wo_cut-off}b and c demonstrate the results with a 6mm rainbow filter and a blade cutoff, respectively.

In the magnified image of fig. \ref{fig:resolution_target_w_wo_cut-off}b, color bands are observable around the numbers and patterns. This phenomenon results from chromatic aberration in our imaging system, despite the use of commercially available achromatic lenses. This limitation restricts the finest resolvable pattern to approximately group 4-6, corresponding to a resolution of $17.5\mu m$ when using the rainbow filter as the cutoff. Conversely, when employing a blade cutoff, the finest resolvable pattern falls between groups 6-5 and 6-6, corresponding to a resolution of approximately $4.6\mu m$. This resolution is primarily limited by the numerical aperture of our optics rather than the cutoff itself.

It is worth noting that while the rainbow filter is susceptible to chromatic aberration, it demonstrates greater robustness to noise as it measures the Hue (color) of the signal. On the other hand, although using a blade cutoff can provide finer resolution, the measurement is more sensitive to background fluctuations, which can introduce noise in brightness measurements. Similar to the choice between double-pass and single-pass systems, there is no universal rule for deciding between a rainbow filter and a blade cutoff; the selection depends on the specific application. In our case, we require high resolution to resolve fine shock wave structures, therefore, we opted for a single-pass Schlieren system combined with a blade cutoff.

\subsection{Schlieren Imaging of Supersonic Jets into Atmospheric}
We investigated supersonic jets produced by a cylindrical nozzle with a radius of 250 $\mu$m and an inlet pressure of 128.9 psi. Under these conditions, the jet reaches a sonic speed at the nozzle exit with a pressure of approximately 78 psi. When the jet is ejected into atmospheric conditions, the pressure ratio between the jet and the ambient environment is greater than one, corresponding to an under-expanding nozzle. Consequently, the jet rapidly expands upon exiting the nozzle, becoming supersonic, and quickly forms a normal shock wave and associated shock structures.

\begin{figure*}[t]
     \centering
     \includegraphics[width=\textwidth]{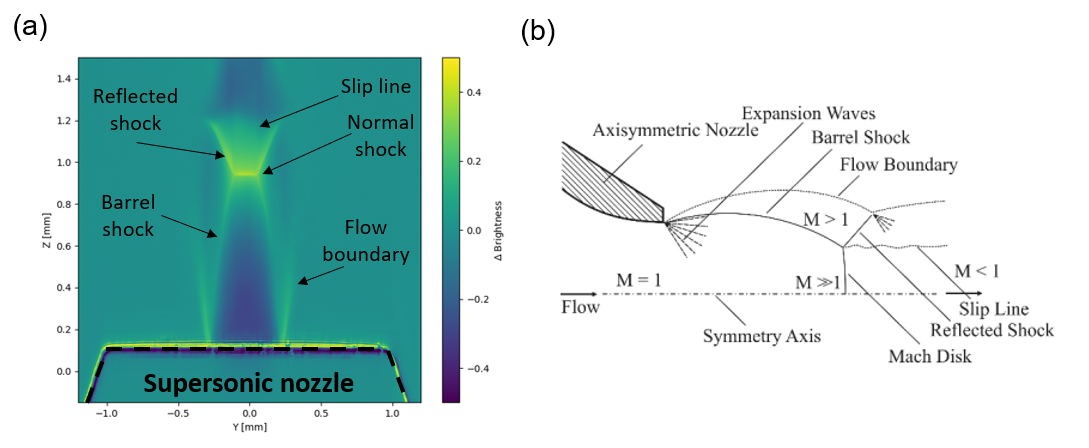}
     \caption{Visualization of a supersonic jet ejected into atmospheric conditions. (a) Schlieren data, where the color scale corresponds to the measured brightness after background subtraction. (b) Schematic representation reproduced from \citep{aerospace2022}.}
     \label{fig:jet_into_air}
 \end{figure*}

Figure.\ref{fig:jet_into_air} presents our Schlieren imaging data alongside a schematic of the jet behavior under these conditions \citep{aerospace2022}. We utilized a vertically oriented blade cutoff, which measures the density gradient along the z-axis. This configuration allows us to observe a change in the sign of the density gradient across the normal shock wave. The high resolution of our imaging system enables visualization of fine structures such as oblique and normal shock waves, as well as the free jet boundaries. Schlieren imaging, by directly measuring the density gradient, produces stronger signals where the density gradient is significant. This characteristic is advantageous for visualizing fine structures in supersonic jets, which typically feature large density gradients. 

\begin{figure*}[t]
     \centering
     \includegraphics[width=\textwidth]{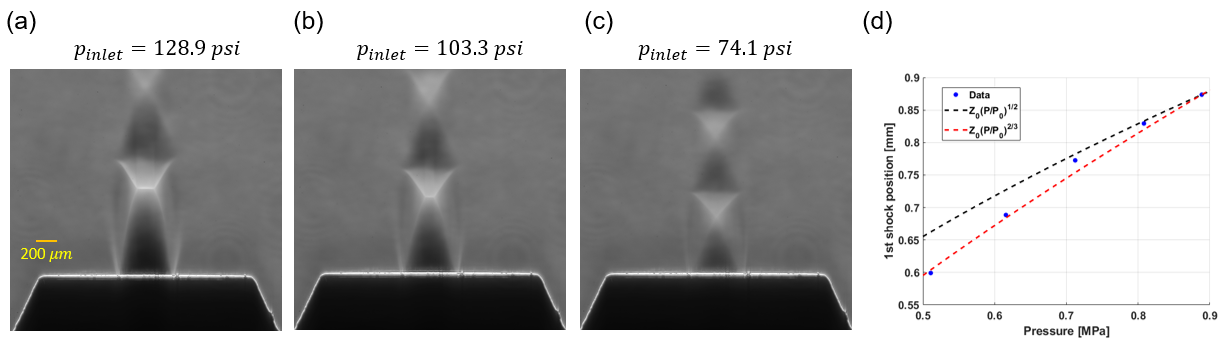}
     \caption{(a)-(c) Visualization of the supersonic jet ejected into atmospheric conditions under different inlet pressure. (d) Position of the first normal shock wave as a function of inlet pressure. The theoretical scaling $Z_{\text{shock}} \propto \sqrt{P_{\text{inlet}}}$ is also shown for comparison. }
     \label{fig:jet_into_air_different_p}
 \end{figure*}

Figure.\ref{fig:jet_into_air_different_p}a-c shows Schlieren images of the supersonic jet injected into the air under different inlet pressures. It is evident that the width of the first normal shock becomes narrower as the inlet pressure decreases. At sufficiently low inlet pressures, the barrel shocks merge, forming the triangular structure (see Fig.\ref{fig:jet_into_air_different_p} (c)). Figure. \ref{fig:jet_into_air_different_p}d compares the position of the first normal shock wave to the theoretical prediction, $Z_{shock}\propto\sqrt{P_{inlet}}$. When the inlet pressure drops below approximately 0.7 MPa, the normal shock wave is no longer visible. In such cases, the position is defined by the port where the barrel shocks converge. This observation may explain the deviation from the theoretical trend at lower pressures.

To validate the quantitative accuracy of our Schlieren imaging system, we compared our experimental data with computer fluid dynamics (CFD) simulations conducted using ANSYS Fluent code. We focused on two key benchmarks: the position of the first shock diamond (normal shock) and the density gradient profile along the nozzle's axis. For the CFD simulations, we used the Reynolds-averaged Navier-Stokes (RANS) solver with a refined cell size of $5\mu m$ in the region of interest. Additionally, finer boundary layers were introduced near the walls to maintain a sufficiently low y-plus parameter, mitigating artifacts in the turbulence model near walls.

For data reconstruction, raw images consist of two sets: one without the jet (background) and another with the jet present. Each set comprised ten shots to minimize system fluctuations. Using a vertically oriented knife blade as the cutoff, the deflection angle was encoded in the brightness channel. We post-processed the images to convert the raw data to HSV color space and extracted the brightness (Value) channel. Calibration results were used to relate brightness to the deflection angle, allowing us to generate a deflection angle map. To deconvolute the integrated signal, we applied Abel inversion using the pyAbel library\citep{pyAbel,pyAbel_BASEX}, reconstructing the density gradient into r-symmetric data. For comparison with the simulation results, we extracted data along the jet's symmetric axis (r=0).

\begin{figure}[h]
    \centering
    \includegraphics[width=1\linewidth]{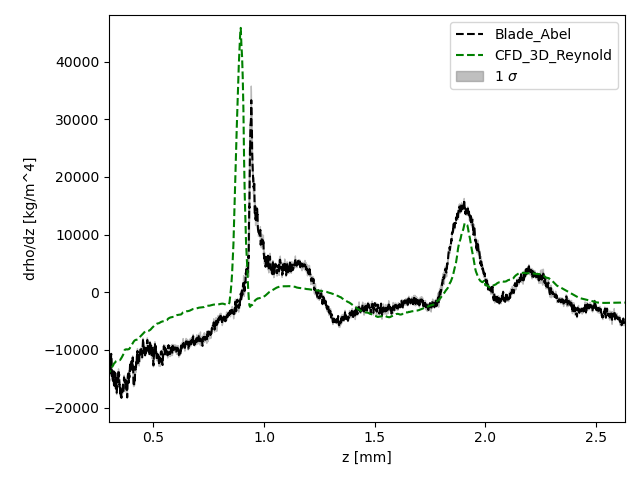}
    \caption{Comparison of density gradient profiles along the symmetry axis between experimental Schlieren data and CFD simulation results for a supersonic jet ejected into atmospheric conditions. The grey-shaded region represents the measurement uncertainty.}
    \label{fig:Jet_into_air_data_Fluent_comparison}
\end{figure}

Figure. \ref{fig:Jet_into_air_data_Fluent_comparison} compares the experimental density gradient with the CFD simulation results. The grey-shaded region represents uncertainty due to measurement fluctuations. Data with $z<0.25\mu m$ is not shown because edge diffraction from the nozzle contaminated the measurements, leading to significant errors in the Abel inversion. The data shows good agreement with the 3D CFD results. The negative density gradient region preceding the first peak indicates rapid jet expansion and acceleration due to the higher exit pressure relative to the ambient pressure. The formation of the normal shock wave at around $z=0.9$ mm above the nozzle exit results in a sharp density discontinuity. After the normal shock wave, the jet expands and contracts periodically, forming shock diamonds, until viscous effects dissipate the flow energy. This explains the second peak in the density gradient at $z\approx1.9$ mm, corresponding to the second shock diamond. The high-resolution qualitative results, together with the good quantitative agreement with CFD simulations, demonstrate the effectiveness of the Schlieren imaging system presented in this paper.

In the following section, we will demonstrate the application of our Schlieren imaging system to measure the density profile of a blade-induced shock wave, a critical application in Laser Wakefield Acceleration (LWA) experiments.
 
\subsection{Shock wave induced by a knife blade obstacle}

\begin{figure}[h]
     \centering
     \includegraphics[width=\linewidth]{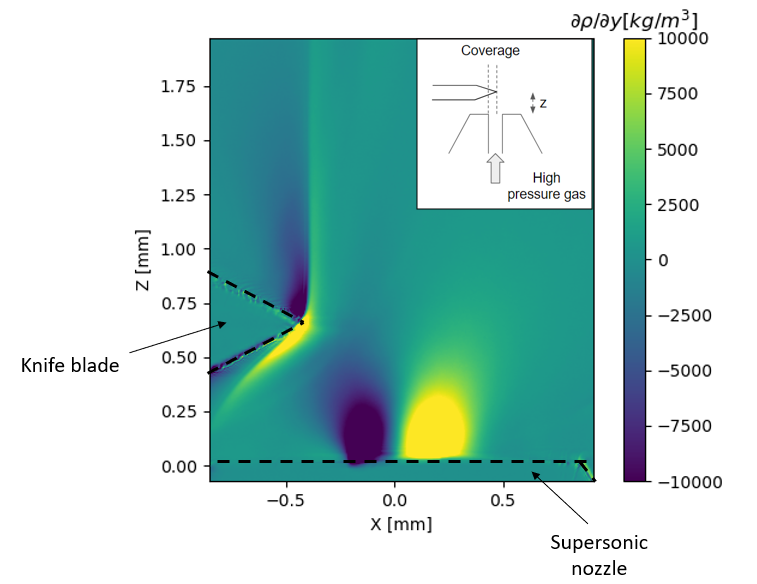}
     \caption{Schlieren image of a blade-induced shock wave, demonstrating the interaction between the supersonic jet and an obstacle placed above the nozzle.}
     \label{fig:blade_induced_shock_wave}
 \end{figure}

By positioning an obstacle above the nozzle, we can induce shock wave formation when the supersonic jet impinges on the obstacle. Figure. \ref{fig:blade_induced_shock_wave} clearly shows a vertical blade-induced shock wave with the blade positioned 700 $\mu$m above the nozzle exit and a coverage ratio of approximately -18\% (the definition of coverage ratio is illustrated in the figure inset). Due to the rapid expansion of the jet after leaving the nozzle, a shock wave can be induced even when the blade tip is outside the nozzle's projection. The improved resolution of our Schlieren system allows us to obtain images of sharp shock waves, enabling a detailed investigation of shock wave properties under various conditions.

\begin{figure}[h]
     \centering
     \includegraphics[width=\linewidth]{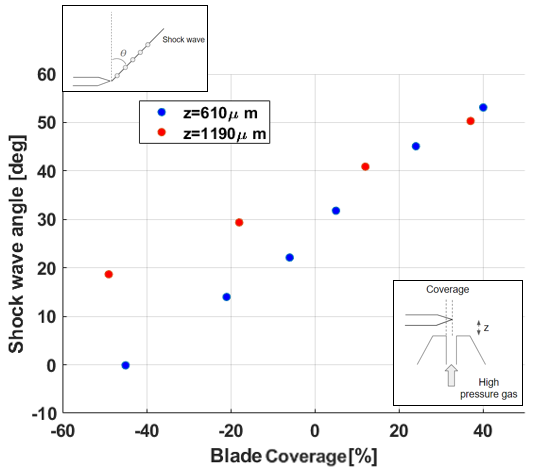}
     \caption{Shock wave angles as a function of blade coverage and height, demonstrating the influence of obstacle configuration on shock wave characteristics.}
     \label{fig:shock_angle_blade_configuration}
 \end{figure}

We investigated the dependence of shock wave angle on blade configurations by scanning different coverage ratios and heights of the blade. Figure. \ref{fig:shock_angle_blade_configuration} presents these results, where shock wave positions at different heights are defined by the location of maximum density gradient. The shock wave angle is obtained by fitting the shock wave position versus height with a linear function. Within our parameter range of interest, the shock waves are predominantly straight, allowing for good linear fits. Generally, the shock wave angle decreases (approaching a vertical shock wave) as the blade coverage decreases. The occurrence of a perfectly vertical shock wave ($\theta = 0^{\circ}$) depends on the blade height. For higher blade positions, the shock wave tends to exhibit a more positive angle.

\begin{figure*}[t]
     \centering
     \includegraphics[width=\textwidth]{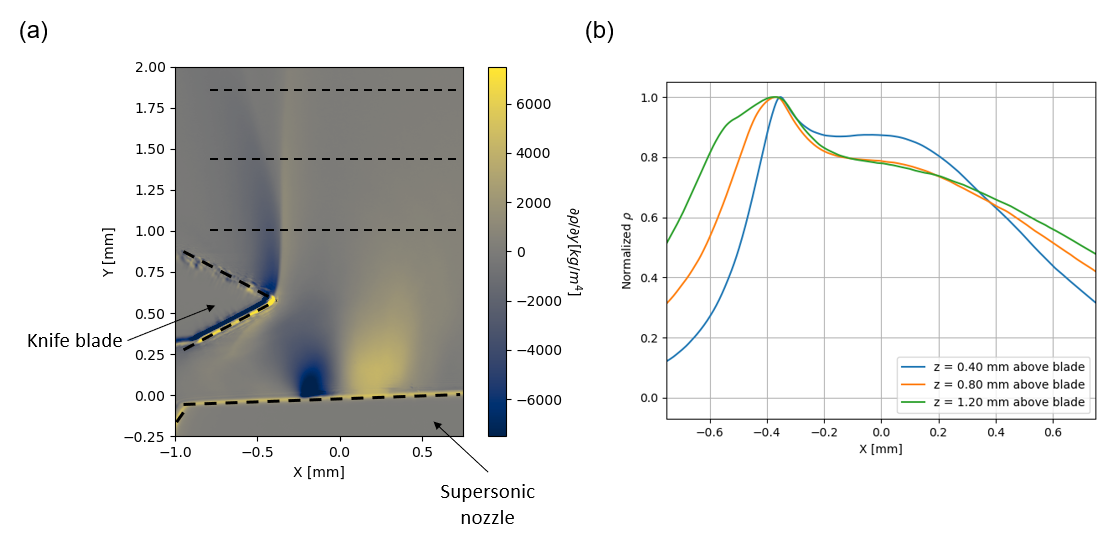}
     \caption{(a) Reconstructed density gradient map of the blade-induced shock wave from the supersonic nozzle. (b) Density profiles across the shock wave at three different heights above the blade.}
     \label{fig:shock_density_profile}
 \end{figure*}

The density profile across the shock wave is of particular interest in applications such as blade-induced shock wave electron injection in laser-plasma wakefield acceleration (LWFA) \citep{bulanov1998injection,schmid2010density}. In these applications, a sharper shock wave is preferable as it leads to a more localized electron injection process, resulting in a narrower energy spread of accelerated electrons. Additionally, the peak density value, plateau length, and plateau density at different heights from the nozzle are critical parameters for optimizing electron acceleration \citep{swanson2017optimization}. Figure. \ref{fig:shock_density_profile} presents an example of our preliminary results showing density profiles across the shock wave for different blade configurations. The data represents an average of ten consecutive shots, each taken after the ambient pressure reached equilibrium. Our Schlieren imaging system successfully captures all the key parameters mentioned above.

The fine resolution of our system also enables the investigation of density variations within the shock wave. In the AnaBHEL experiment, a density profile of the form $\rho(x) = n_0(1 + ae^{-x/D})^2$ was proposed for realizing a specific trajectory of a flying plasma mirror traversing such a plasma background \citep{pisin2020mirrorTrajectory}. Two critical aspects of the gaseous target are the form of the density profile and the parameter $D$, which determines the temperature of the analog black hole (smaller $D$ corresponds to higher temperature).

\begin{figure*}
     \centering
     \includegraphics[width=\linewidth]{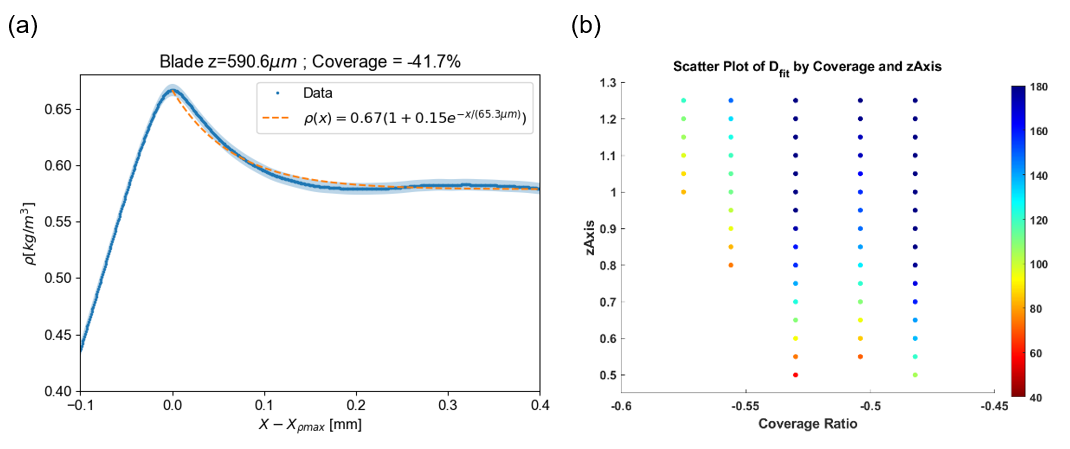}
     \caption{(a) Density profile at z=0.4mm from the blade tip, fitted with the function $\rho(x) = \rho_0(1 + be^{-x/D})$, which represents the desired density profile for the AnaBHEL experiment. (b) Parameter scan of $D$ for a blade positioned 590.6 $\mu$m above the nozzle with varying coverage.}
     \label{fig:density_profile_fitted_one_plus_exp}
 \end{figure*}

We found that the density profile across a shock wave closely approximates the desired density form. Figure. \ref{fig:density_profile_fitted_one_plus_exp}a shows a section of the density profile fitted with the one-plus-exponential formula, yielding $D \approx 65.3 \mu$m. The shaded blue region represents the error bar, primarily due to fluctuations during measurement. Although this value of D is still far from the proposed $0.5 \mu$m in \citep{pisin2020mirrorTrajectory}, it demonstrates the feasibility of using blade-induced shock waves for tailoring desired density profiles.

Figure. \ref{fig:density_profile_fitted_one_plus_exp}b presents a map of the fitted parameter $D$ from density profiles at different heights from the nozzle under various blade coverage. This map can serve as a guideline for blade adjustment in experiments to optimize key parameters. While a detailed discussion of shock wave properties is beyond the scope of this paper, a more comprehensive analysis will be reported in a separate publication. Here, we emphasize that the improved resolution of our Schlieren imaging system enables the investigation of fine structures in supersonic jets and shock waves from sub-millimeter nozzles.

\section{Conclusions}
In this study, we have presented a single-pass quantitative Schlieren imaging system featuring a novel design that allows for independent adjustment of sensitivity and resolution. Our system achieves an optical resolution of $4.6\mu m$, enabling detailed investigation of fine structures in supersonic jets emanating from sub-millimeter nozzles. This resolution can be further enhanced by switching the imaging optics and relay lenses or employing optics with larger numerical apertures, such as relay lenses with increased diameters, and shorter focal lengths or higher magnification objectives. To address the averaging effect caused by the integration of density gradients along the line of sight, future work could incorporate 3D tomography by combining the current system with a rotary stage, allowing for multi-angle scanning. This would mitigate line-of-sight averaging and enhance the reconstruction of 3D structures.

We have conducted preliminary analyses of shock wave behavior resulting from the interaction between a supersonic jet and an impinging blade. These investigations included studying the relationship between shock wave angle and blade configuration, as well as mapping the density profile across the shock wave. These results demonstrate the system's capability to capture and quantify complex fluid dynamic phenomena with high resolution. Beyond fluid dynamics, our Schlieren system shows significant potential for applications in laser-plasma experiments, such as laser wakefield acceleration and the Analog Black Hole Evaporation via Laser (AnaBHEL) experiment. The versatility and precision of the system suggest it could be a valuable tool for exploring hydrodynamics phenomena across micrometer to millimeter scales.

\section{Acknowledgement}
We thank the National Center for High-performance Computing (NCHC) in Taiwan, R.O.C. for providing computational and storage resources.

\clearpage
\balance
\bibliography{main}

\end{document}